%% file: proc1.tex
\documentclass[12pt]{article}
\usepackage{graphicx}

\newcommand\pubnumber{}
\newcommand\pubdate{\today}

\def\napoli{Centro Brasileiro de Pesquisas F\'{\i}sicas, Rio de Janeiro, Brazil}

\def\Title#1{\begin{center} {\Large #1 } \end{center}}
\def\Author#1{\begin{center}{ \sc #1} \end{center}}
\def\Address#1{\begin{center}{ \it #1} \end{center}}

\newcommand\pubblock{\rightline{\begin{tabular}{l} \pubnumber\\
         \pubdate  \end{tabular}}}
\newenvironment{Abstract}{\begin{quotation}  }{\end{quotation}}
\newenvironment{Presented}{\begin{quotation} \begin{center} 
             PRESENTED AT\end{center}\bigskip 
      \begin{center}\begin{large}}{\end{large}\end{center} \end{quotation}}
\def\Acknowledgements{\bigskip  \bigskip \begin{center} \begin{large}
             \bf ACKNOWLEDGEMENTS \end{large}\end{center}}

\input econfmacros.tex

\begin{document}
\begin{titlepage}
\pubblock

\vfill
\Title{CP violation in charm decays}
\vfill
\def\support{\footnote{On behalf of the LHCb collaboration.}}
\Author{ Alberto C. dos Reis,\support}
\Address{\napoli}
\vfill
\begin{Abstract}
This note presents LHCb results on CP violation searches performed on 
charm decays that have been released after FPCP2012.
\end{Abstract}
\vfill
\begin{Presented}
Flavor Physics and CP Violation\\
Buzios, Brazil,  May 19--24, 2013
\end{Presented}
\vfill
\end{titlepage}
\def\thefootnote{\fnsymbol{footnote}}
\setcounter{footnote}{0}

\section{Introduction}

In the  Standard Model (SM) {\em CP} violation (CPV) in the charm sector is 
restricted to Cabibbo-suppressed decays and is predicted to be very small.
Due to the smallness of the SM expectations, CPV in charm has been seen as a 
portal to new physics (NP). Indeed, until recently it was often assumed that the
observation of {\em CP} asymmetries in $D$ meson decays at the $\mathcal{O}(10^{-2})$ 
level would be an indication of new sources of CPV. SM predictions, however, suffer 
from the large uncertainty on the magnitude of penguin amplitudes.
According to recent estimates, it is conceivable that {\em CP} asymmetries at 
this level could be accommodated within the
SM \cite{grossman,brod,gronau}.

From the experimental side, the sensitivity of CPV searches has increased 
dramatically in the past few years, especially with the advent of the large LHCb 
data sets. {\em CP} asymmetries at the $\mathcal{O}(10^{-2})$ level in $D$ meson decays  
seem now to be excluded. With errors reaching the level of a few per mille, the 
current CPV searches probe a regime where {\em CP} asymmetries  would be 
consistent with the SM expectations. The interpretation of an eventual
observation of {\em CP} asymmetries in charm would not be straightforward.

This note collects recent LHCb results on CPV searches, all based on the full
2011 data set (1.0 fb$^{-1}$ at $\sqrt{s}=$ 7 TeV). An update of the measurement of
the difference in time-integrated {\em CP} asymmetry between $D^0\to K^-K^+$ and 
$D^0\to \pi^-\pi^+$,
$\Delta A_{CP} \equiv A_{CP}(D^0\to K^-K^+) - A_{CP}(D^0\to \pi^-\pi^+)$ 
\cite{conf3}, is presented, along with results from an independent
measurement of the same observable using $D^0$ from semileptonic $b$-hadron decays. 
Results from CPV searches using charged $D$ mesons are also reviewed. 

\section{Update on $\Delta A_{CP}$ from prompt $D^{*+}$}

\subsection{Analysis strategy and sample selection}

As in the previous measurement \cite{dacp11}, the flavor of the initial
state is determined by the charge of the soft pion ($\pi^+_s$)  in the
decays $D^{*+} \to D^0 \pi^+_s$ and $D^{*-} \to \overline{D}^0 \pi^-_s$.

The decay-time-dependent {\em CP} asymmetry $A_{CP}(f;t)$ for the decays
$D^0 \to f$ is defined as

\begin{equation}
A_{CP}(f;t)=\frac{\Gamma(D^0(t)\to f)-\Gamma(\overline{D}^0(t)\to f)}
{\Gamma(D^0(t)\to f)+\Gamma(\overline{D}^0(t)\to f)},
\end{equation}
where $f$ is a {\em CP} eigenstate, $f=K^-K^+,\pi^-\pi^+$. One can express $A_{CP}(f;t)$
as a sum of two terms: a direct, time-independent component, associated with CPV in the 
decay amplitudes and dependent on the final state; and
an indirect, time-dependent component, universal to a good approximation, 
associated with CPV in mixing or in the interference between mixing and  decay.
The time-integrated asymmetry $A_{CP}(f)$ may be written to first order as

\begin{equation}
A_{CP}(f) = a^{\mathrm{dir}}_{CP} + \frac{\langle t \rangle}{\tau}a^{\mathrm{ind}}_{CP},
\end{equation}
where $\langle t \rangle$ is the average decay time in the reconstructed sample and
$\tau$ is the average $D^0$ lifetime. 

The difference between {\em CP} asymmetries for the $D^0 \to K^-K^+$ and $D^0 \to \pi^-\pi^+$
is thus

\begin{equation}
\Delta A_{CP} = a^{\mathrm{dir}}_{CP}(K^-K^+) -a^{\mathrm{dir}}_{CP}(\pi^-\pi^+) +
\frac{\Delta \langle t \rangle}{\tau}a^{\mathrm{ind}}_{CP}.
\end{equation}

In this analysis we measure $\Delta \langle t \rangle/\tau=[11.19\pm 0.13\pm 0.17]$\%.
Indirect {\em CP} violation is a small contribution to $\Delta A_{CP}$.

The raw asymmetry for tagged $D^0$ decays to a final state $f$ is

\begin{equation}
A_{\mathrm{raw}}(f)=\frac{N(D^{*+}\to D^0(f)\pi^+_s)-
N(D^{*-}\to \overline{D}^0(f)\pi^-_s)}
{N(D^{*+}\to D^0(t)\pi^+_s)+
N(D^{*-}\to \overline{D}^0(t)\pi^-_s)},
\end{equation}
where $N$ is the number of reconstructed events after background subtraction.

The raw charge asymmetry has a component due to detector, $A_D$, and production, 
$A_P$, effects. If these effects are small, the raw asymmetry is, to first order, 

\begin{equation}
A_{\mathrm{raw}}(f)= A_{CP}(f) + A_D(f) + A_D(\pi^+_s) + A_P(D^{*+}).
\end{equation}

Since the two modes are self-conjugate final states, there is no detection asymmetry:
$A_D(K^-K^+)=A_D(\pi^-\pi^+)=0$. $A_D(\pi^+_s)$ and $A_P(D^{*+})$ are independent of
the final state. These terms cancel, to first order, in the difference
$A_{\mathrm{raw}}(K^-K^+)-A_{\mathrm{raw}}(\pi^-\pi^+)$:

\begin{equation}
\Delta A_{CP} = A_{\mathrm{raw}}(K^-K^+) - A_{\mathrm{raw}}(\pi^-\pi^+)
\end{equation}

The following presents a terse summary of the sample selection,
 which is fully described in Ref. \cite{conf3}. 
The sample is selected imposing requirements on the kinematic properties and on the
decay time of the $D^0$ candidate. The  $D^0$ decay products are required to form a
displaced vertex with good fit quality, pointing back to the primary vertex. It is also
required that the $D^0$ daughter tracks have good fit quality, transverse momentum 
above a minimum value and momentum directions that do not point to the primary vertex. 
The RICH system is used to
distinguish between kaon and pions. Fiducial requirements are imposed to exclude 
kinematic regions with large charge asymmetry in the soft pion detection efficiency.
The candidates must pass the specific high level software trigger.
The $D^{*+}$ yields are $2.24 \times 10^6$ for $D^0 \to K^-K^+$ decays, and
$0.69\times 10^6$ for $D^0 \to \pi^-\pi^+$ decays.

\subsection{Differences from previous analysis}

There are some changes to the reconstruction and analysis procedure
with respect to the previous measurement:

\begin{itemize}
\item full 2011 data set (1.0 fb$^{-1}$ at 7 TeV);
\item improved reconstruction, with a more accurate alignment and calibration;
\item weighting procedure to account for small remaining differences in the kinematics
of the $D^{*+}\to  D^0(K^-K^+)\pi_s^+$ and $D^{*+}\to  D^0(\pi^-\pi^+)\pi_s^+$ decays;
\item better mass resolutions due to a constrained fit that forces the 
soft pion to come from the primary vertex.
\end{itemize}

\subsection{Fit model and results}

For each $D^0$ final state the data are divided into four disjoint samples 
according to magnet polarity and hardware trigger decision.  The raw asymmetry 
is determined, for each final state, by a simultaneous fit of the 
$\delta m\equiv m(h^-h^+\pi^+) - m(h^-h^+) - m(\pi^+)$ ($h=K,\pi$) distributions 
from the four sub-samples. The $\delta m$ spectrum is obtained accepting $D^0$
candidates in the mass interval 1.844$<m(h^-h^+)<$1.884 MeV/$c^2$

\begin{figure}[htb]
\centering
\includegraphics[width=7cm]{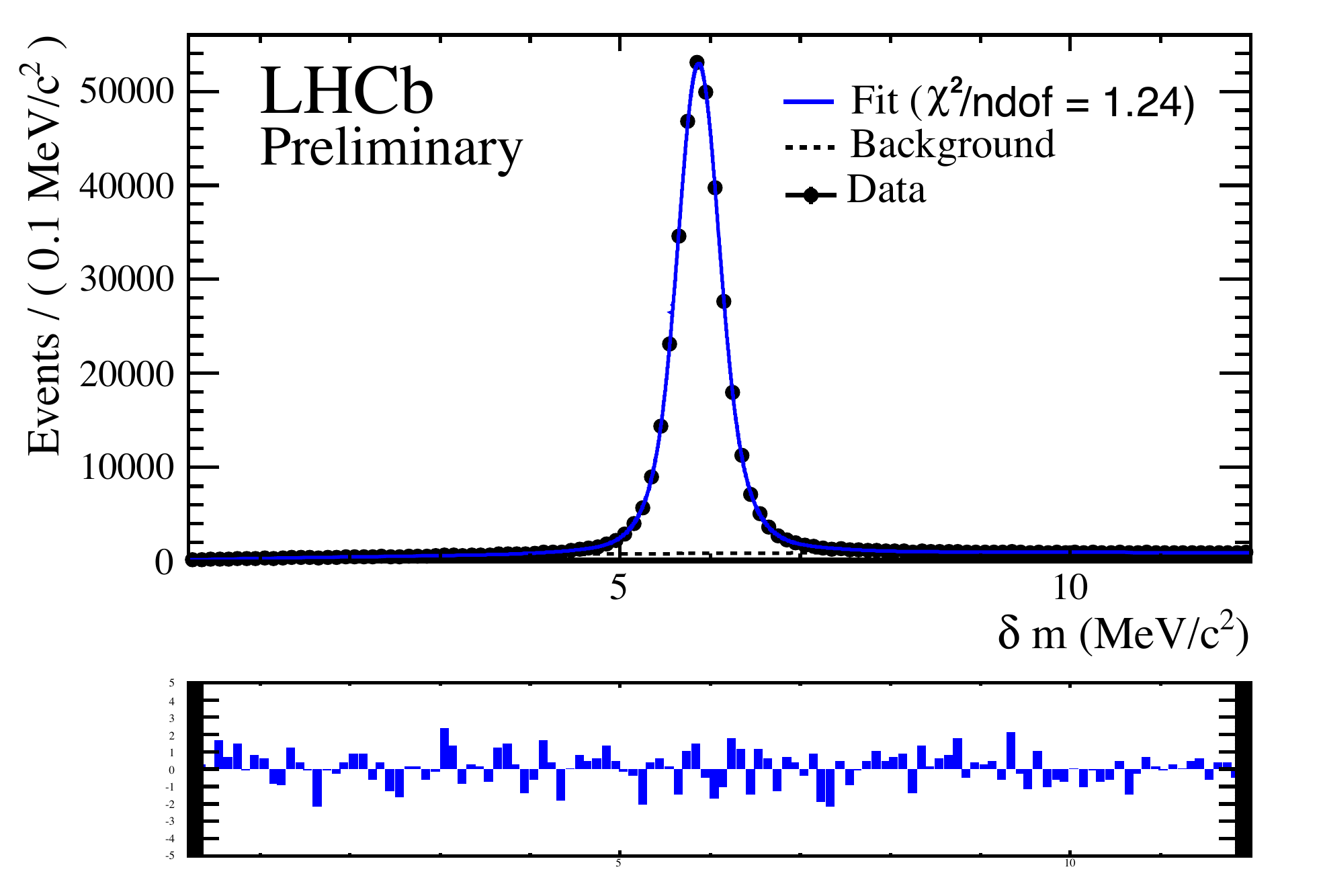}
\includegraphics[width=7cm]{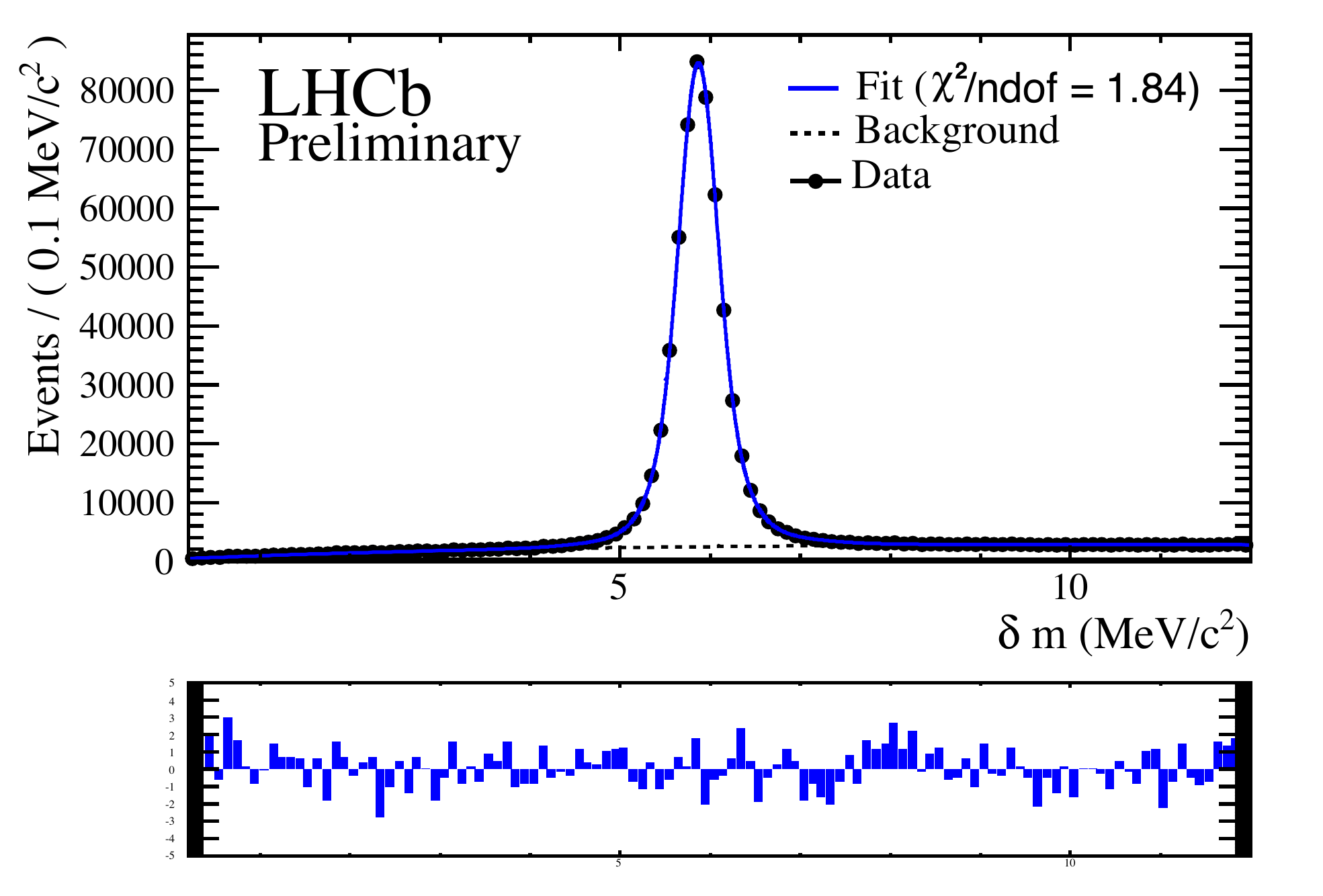}

\includegraphics[width=7cm]{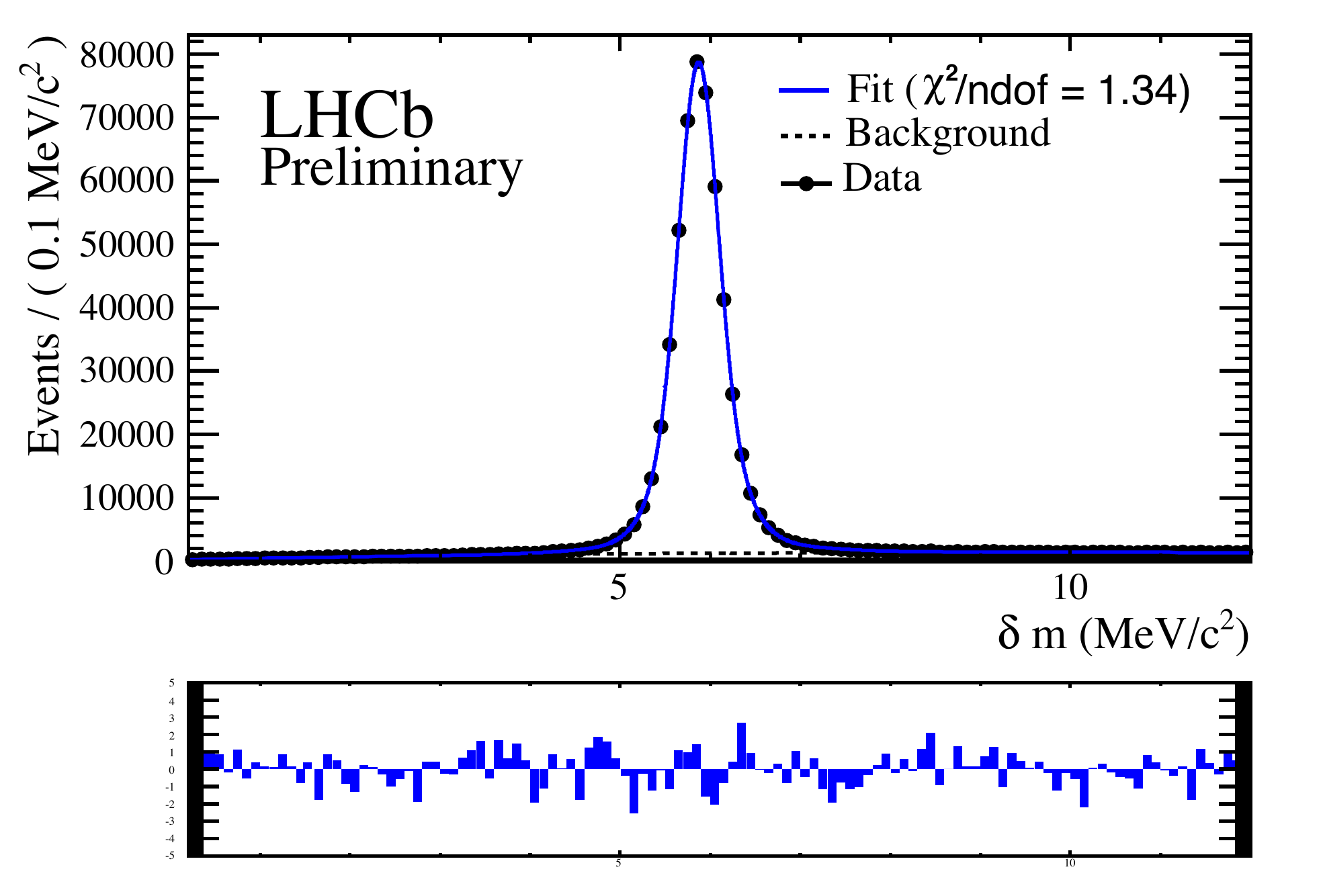}
\includegraphics[width=7cm]{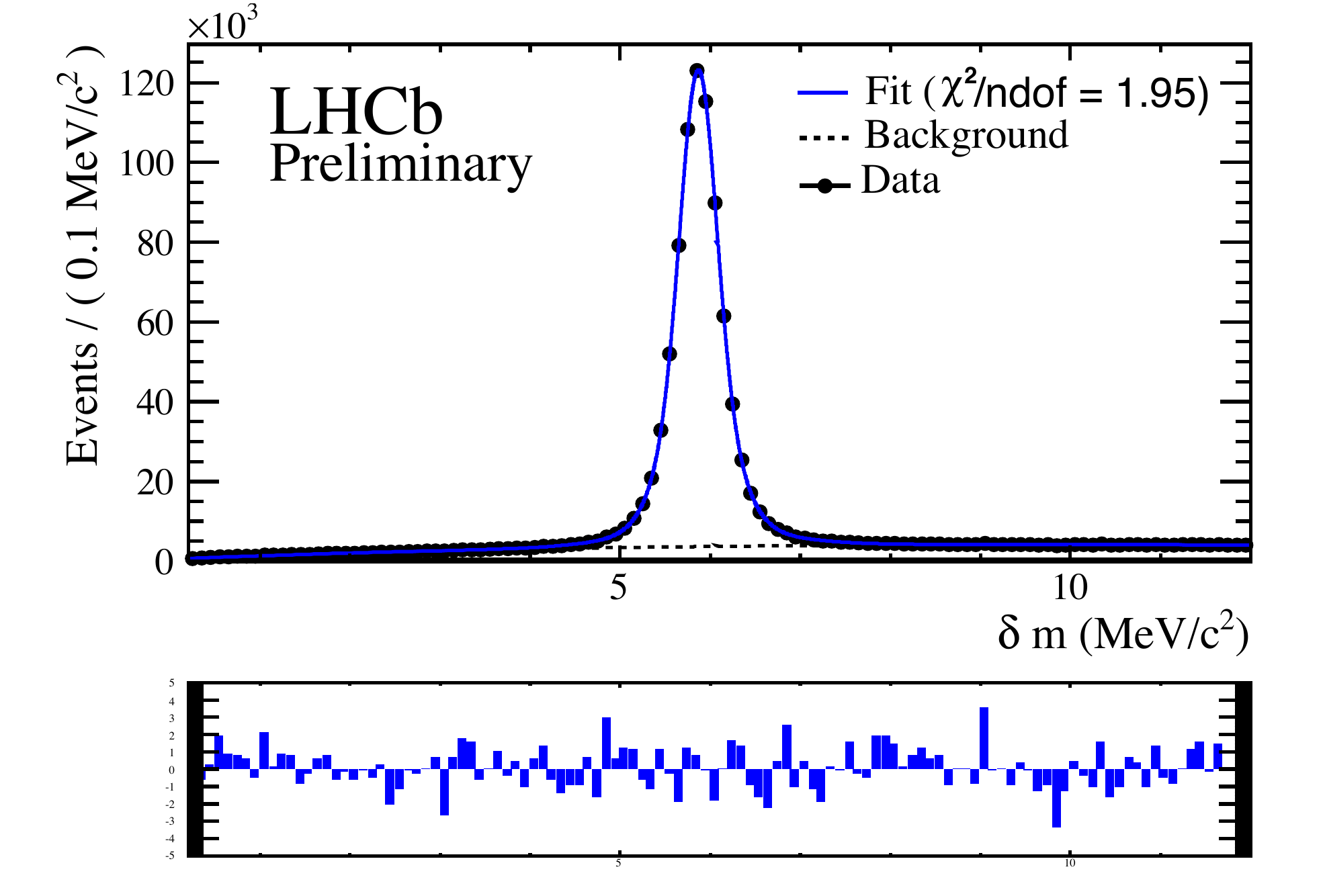}
\caption{Fits to the $\delta m$ spectra from $D^{*+}\to D^0(K^-K^+)\pi_s^+$.
Candidates are divided into four independent  sub-samples according to 
magnet polarity and hardware trigger decision. The normalized residuals 
are shown below the fits. The fit procedure is described in the text.}
\label{fig:dacp1}
\end{figure}

\begin{figure}[htb]
\centering
\includegraphics[width=7cm]{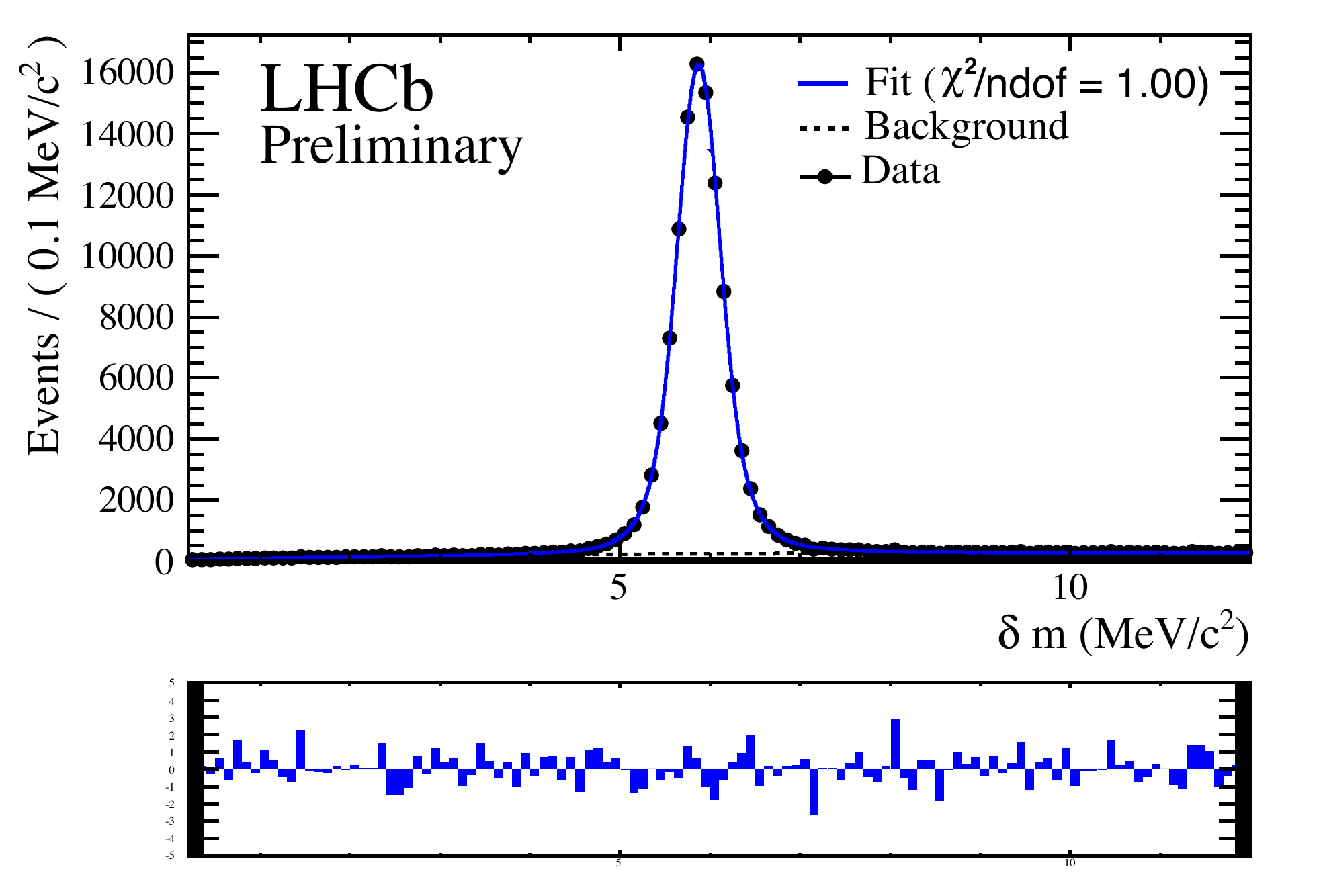}
\includegraphics[width=7cm]{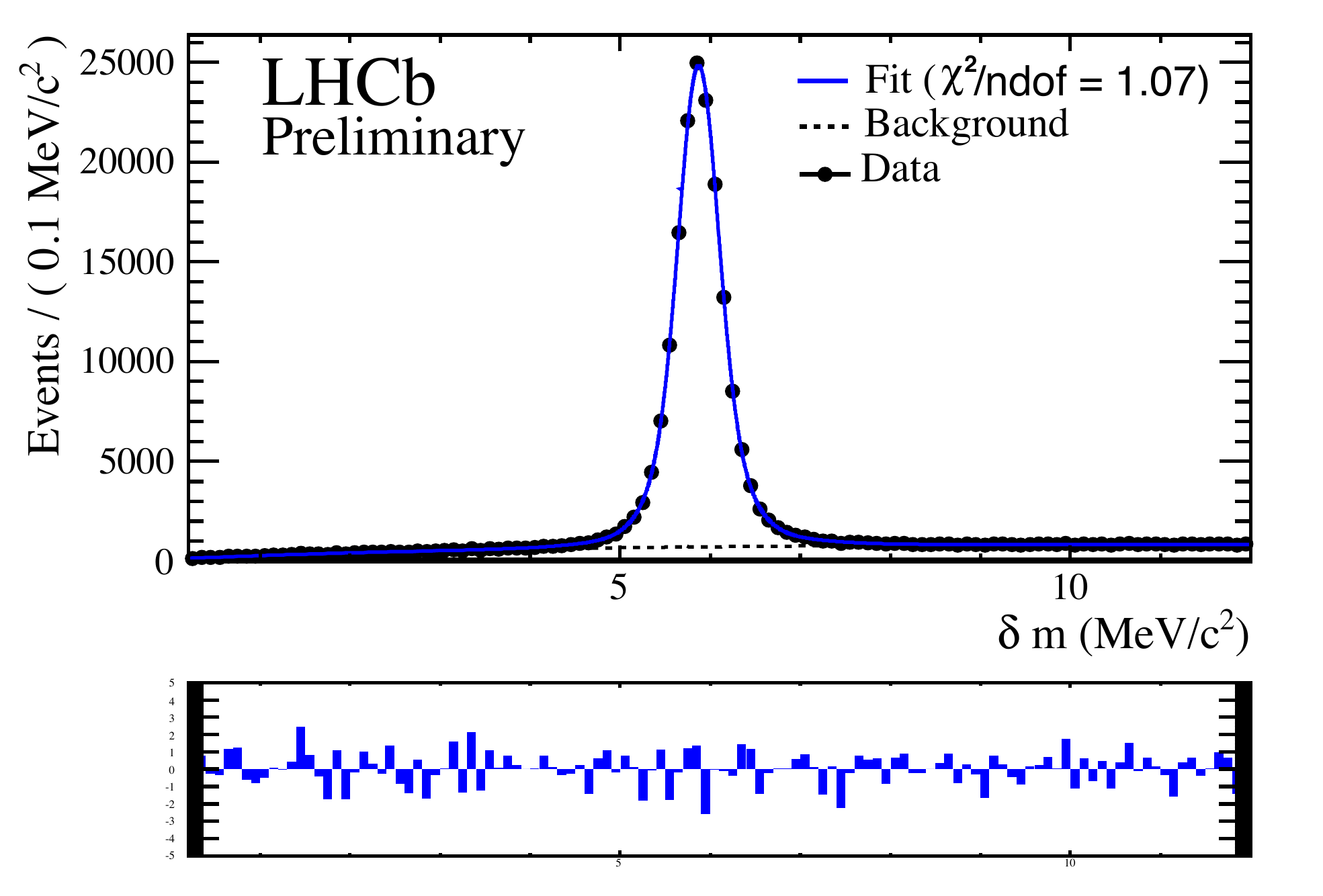}

\includegraphics[width=7cm]{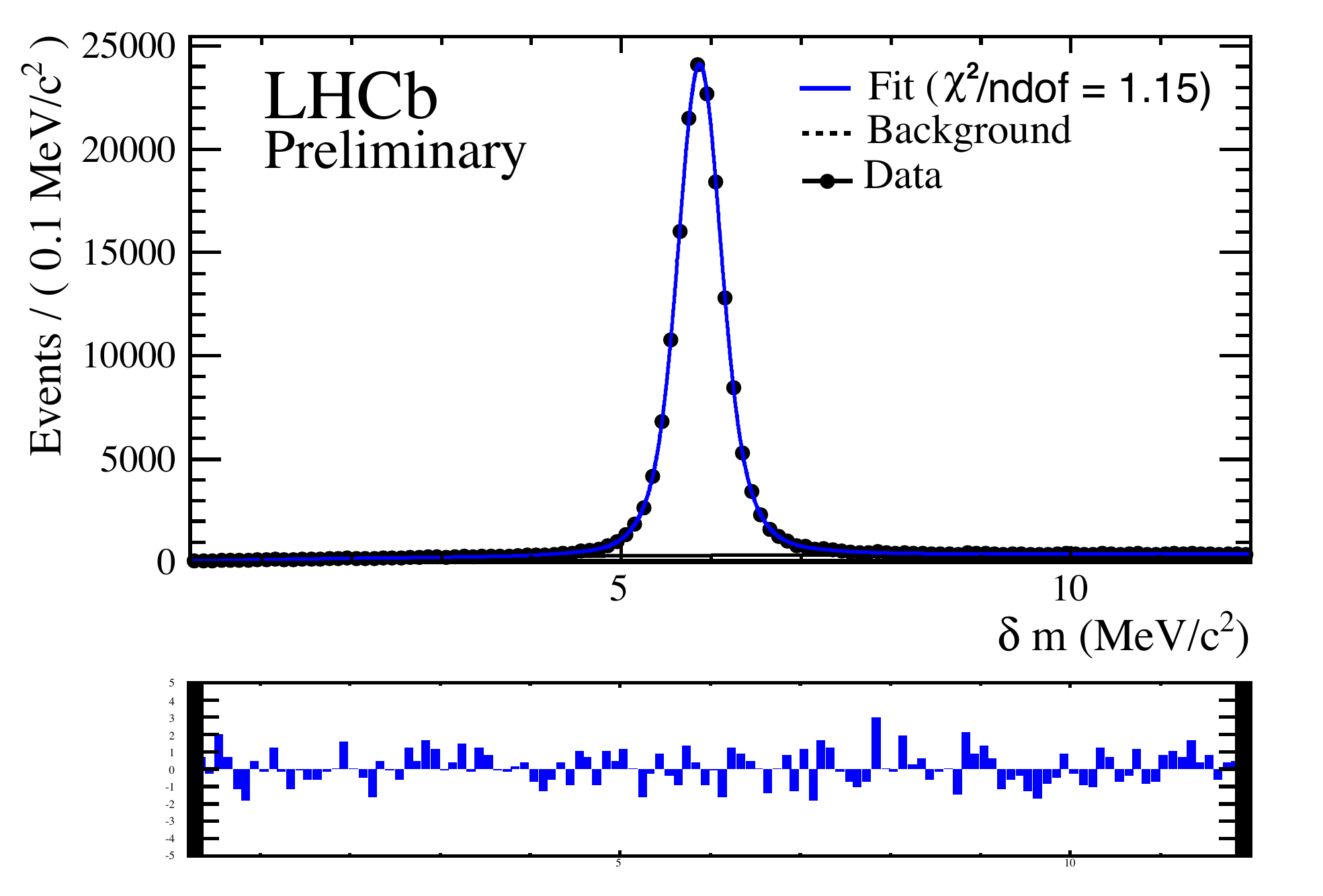}
\includegraphics[width=7cm]{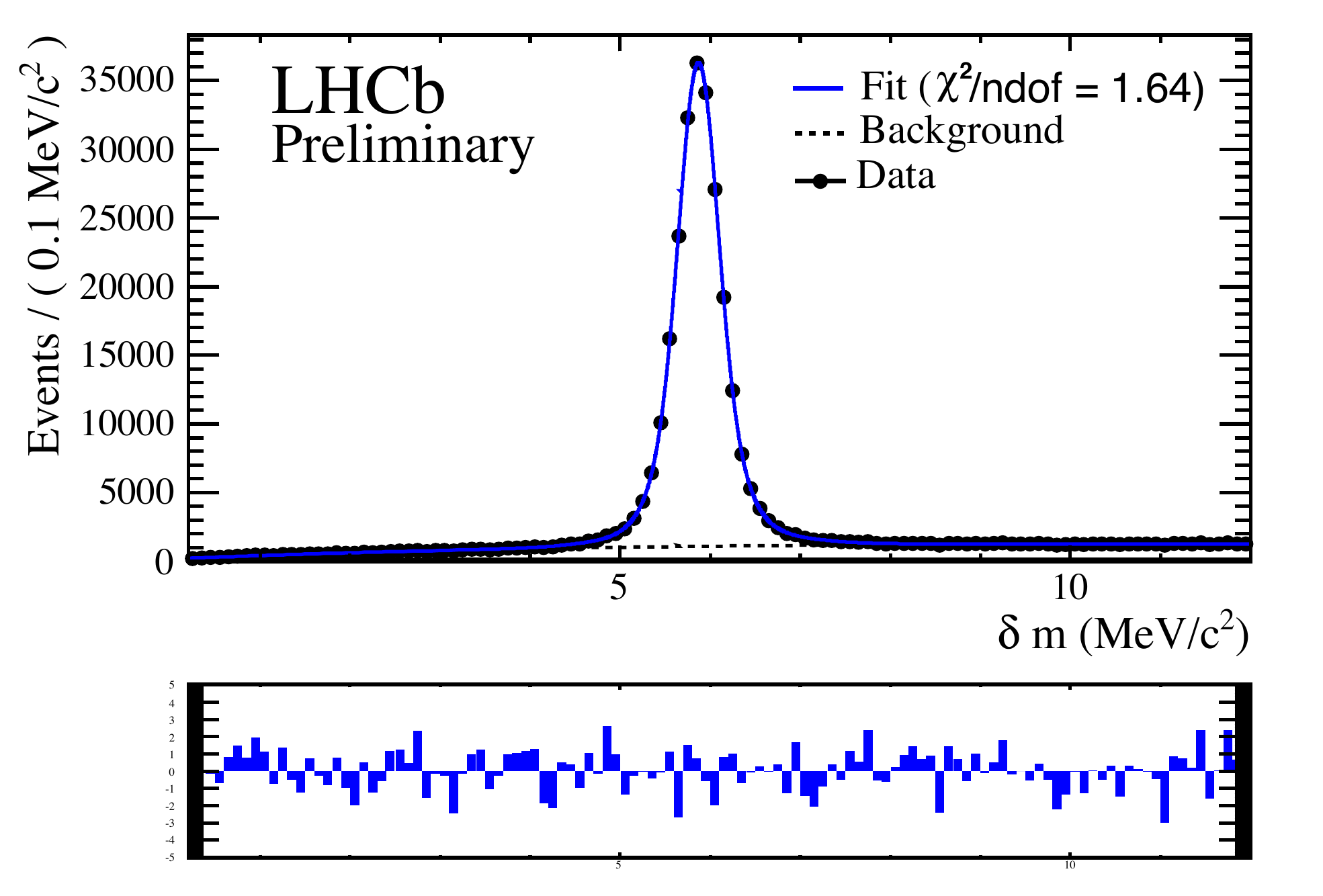}
\caption{Fits to the $\delta m$ spectra from $D^{*+}\to D^0(\pi^-\pi^+)\pi_s^+$ with
 normalized residuals. Candidates are divided into four independent  sub-samples according to 
magnet polarity and hardware trigger decision. The fit procedure is described in the text.}
\label{fig:dacp2}
\end{figure}

The signal is described by a sum of three Gaussian functions with a common mean convolved
with an asymmetric function to improve the description of the tails.  The background is
described by the empirical function
$B(\delta m)\propto [\delta m - m_0]^a
e^{-b(\delta m - m_0)}$. The parameters $a$, $b$ and $m_0$ describe the shape of the function.

For each subsample --- magnet polarity and hardware trigger category 
--- $\Delta A_{CP}$ is computed.
The combined value is computed as a weighted average across the sub-samples. 
Systematic uncertainties 
are assigned by loosening the fiducial requirement on the soft pion; by estimating the impact of
a potential
background from mis-reconstructed $D^{*+}$; by evaluating the asymmetry using sideband 
subtraction instead of a fit; 
by comparing with the result obtained with no kinematic weighting; and by excluding events
in which the soft pion has a large impact parameter with respect to the primary vertex. The latter
is the dominant source of systematic uncertainty. 

The final result is

\begin{equation}
\Delta A_{CP} = (-0.34\pm0.15(\mathrm{stat.})\pm0.10(\mathrm{syst.}))\%
\end{equation}

\section{$\Delta A_{CP}$ from semileptonic $b$-hadron decays to $D^0\mu^-X$}

\subsection{Analysis strategy}

An independent measurement of $\Delta A_{CP}$ was performed using $D^0$ from
semileptonic $b$-hadron decays to $D^0\mu^-X$ \cite{semilep}.
 The initial flavor of the  $D^0$ 
is tagged by the charge of the muon: a positive muon is associated with a
$\overline{D}^0$, and a negative muon with a $D^0$ meson. The $X$ denotes
any other particle(s) that are not reconstructed.

The raw charge asymmetry is written in terms of the $D^0$ decay rate $\Gamma$,
the muon detection efficiency $\varepsilon$, and the  $D^0$ production rate
in semileptonic $b$-hadron decays, $\mathcal{P}$,

\begin{equation}
A_{\mathrm{raw}} = \frac{\Gamma(D^0)\varepsilon(\mu^-)\mathcal{P}(D^0)-
\Gamma(\overline{D}^0)\varepsilon(\mu^+)\mathcal{P}(\overline{D}^0)}
{\Gamma(D^0)\varepsilon(\mu^-)\mathcal{P}(D^0)+
\Gamma(\overline{D}^0)\varepsilon(\mu^+)\mathcal{P}(\overline{D}^0)}.
\end{equation}

The raw asymmetry can be written to first order as

\begin{equation}
A_{\mathrm{raw}} \simeq A_{CP} + A^{\mu}_D + A_{\mathcal{P}}^B,
\end{equation}
with the definitions 

\[
A_{CP} = \frac{\Gamma(D^0)-\Gamma(\overline{D}^0)}
{\Gamma(D^0)+\Gamma(\overline{D}^0)}, \hskip .3cm 
A^{\mu}_D = \frac{\varepsilon(\mu^-)-\varepsilon(\mu^+)}
{\varepsilon(\mu^-)+\varepsilon(\mu^+)}, \hskip .3cm 
A_{\mathcal{P}}^B = \frac{\mathcal{P}(D^0)-\mathcal{P}(\overline{D}^0)}
{\mathcal{P}(D^0)+\mathcal{P}(\overline{D}^0)}.
\]

As in the analysis with prompt $D^{*+}$, the difference between the raw 
asymmetries measured in $D^0\to \pi^-\pi^+$ and $D^0\to K^-K^+$ decays cancels
the detection and production asymmetries,

\begin{equation}
\Delta A_{CP}= A_{\mathrm{raw}}(K^-K^+) - A_{\mathrm{raw}}(\pi^-\pi^+) \simeq
A_{CP}(K^-K^+) - A_{CP}(\pi^-\pi^+).
\label{dacpsemi}
\end{equation}

The cancellation requires the
kinematic distributions of the muon and the $b-$hadron to be the same for
both $D^0$ final states. This is ensured by a weighting procedure which
equalises the kinematic distributions.

In this analysis the ratio between the difference of the $D^0$ and  $\overline{D}^0$  
mean decay time and the world average $D^0$ lifetime is

\[
\frac{\Delta \langle t \rangle}{\tau} =  (1.8\pm0.2\pm0.7)\%,
\]
making the contribution of indirect {\em CP} violation in eq. 3 marginal.

\subsection{Data set and selection}

As for the prompt $D^{*+}$ sample,
candidates are required to be accepted by a specific trigger decision. 
In the hardware stage about
87\% of candidates in the final selection are accepted by the hardware
trigger based only in the muon system, 3\% are accepted by the hadronic 
calorimeter only, and the remaining 10\% by both systems. In the software stage
candidates are selected by either a single muon trigger or by a topological
trigger \cite{trig}. The offline selection imposes requirements on track and vertex fit 
quality, on impact parameter for each track with respect to the primary vertex
and on transverse momentum of the muon and the $D^0$ daughters. Requirements
on the $D^0$ decay topology are minimal in order to have similar decay-time 
acceptance for both $D^0$ final states. Particle identification is required
for all tracks. A detailed description of the sample selection is found in
\cite{semilep}.

\begin{figure}[htb]
\centering
\includegraphics[width=7cm]{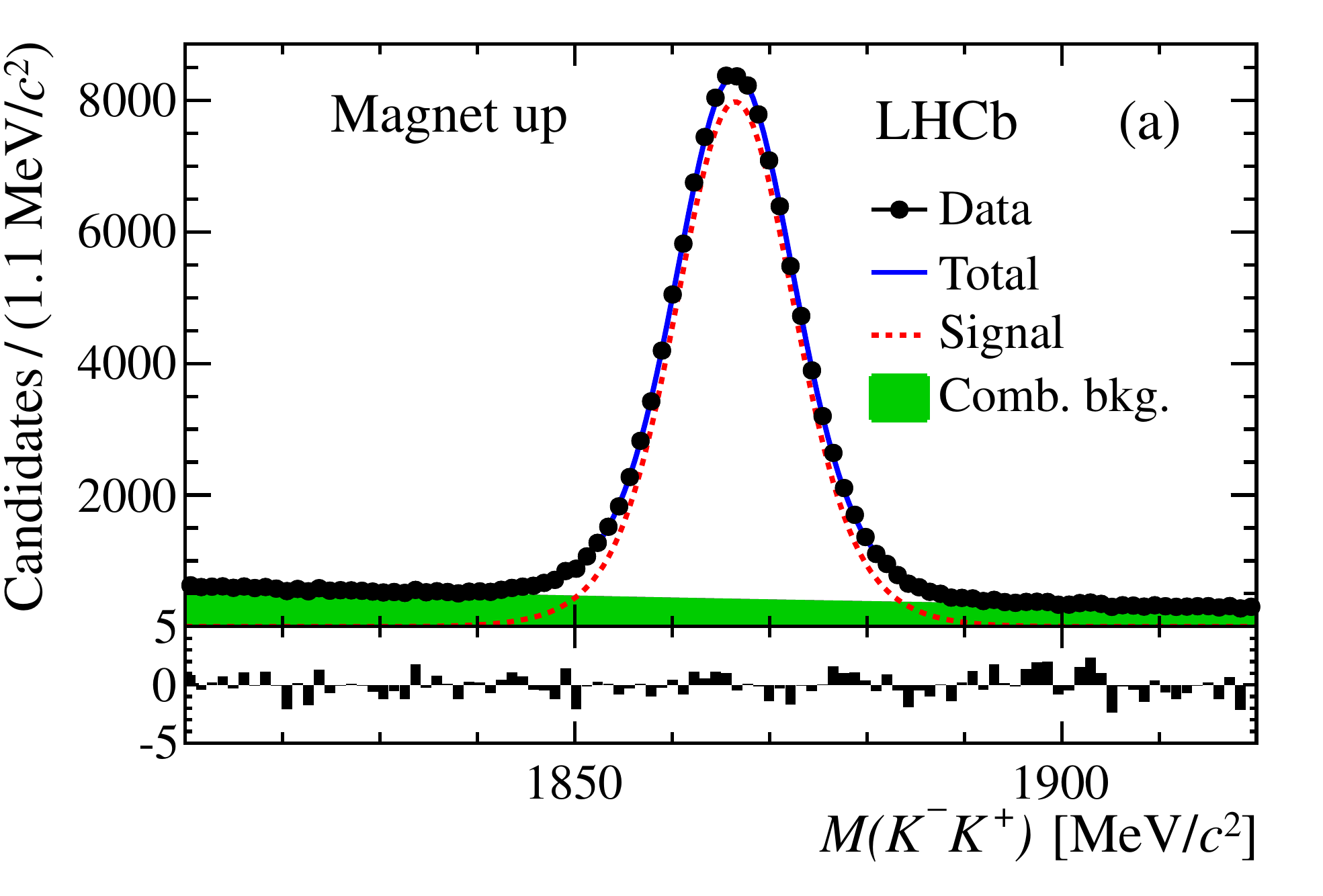}
\includegraphics[width=7cm]{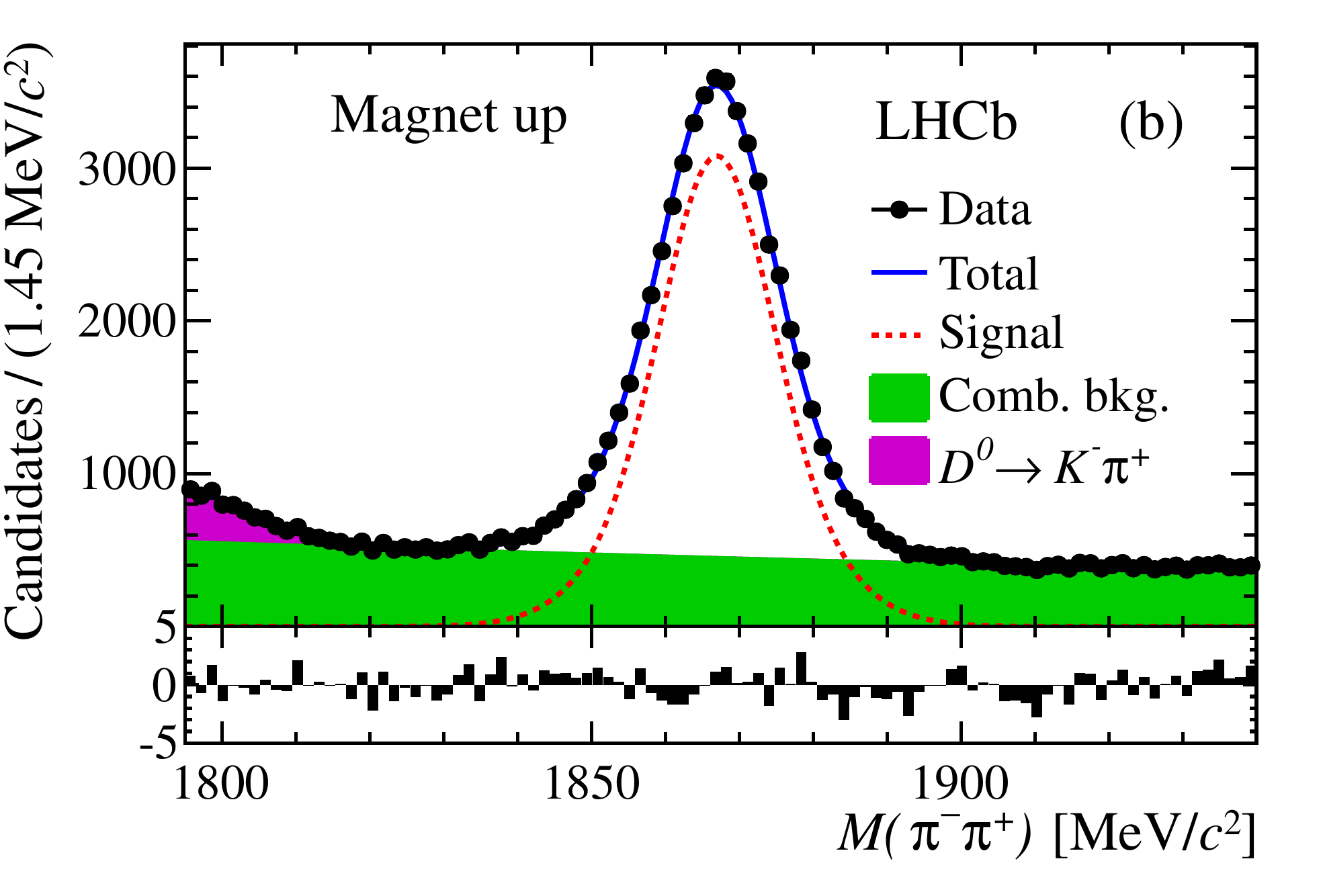}

\includegraphics[width=7cm]{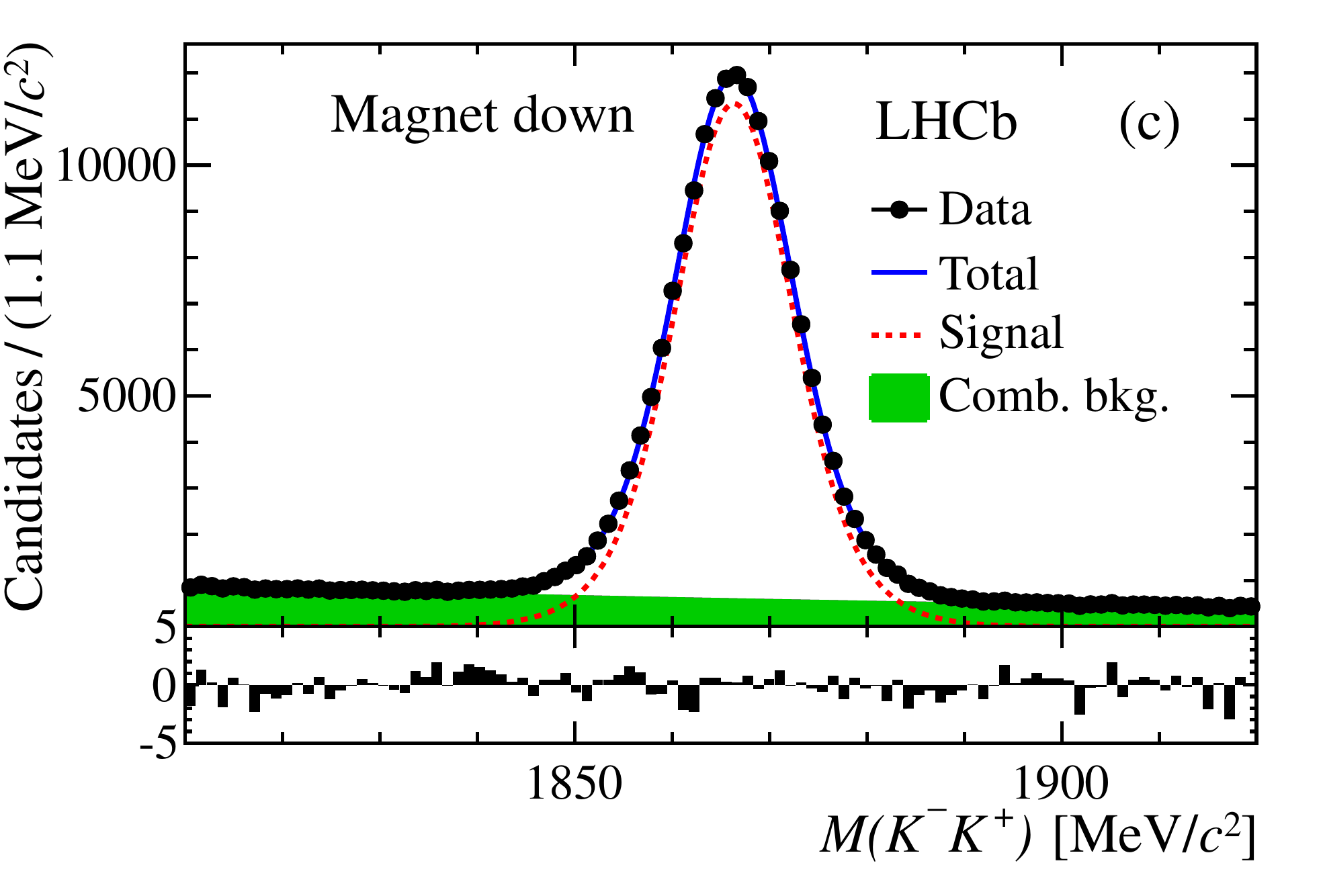}
\includegraphics[width=7cm]{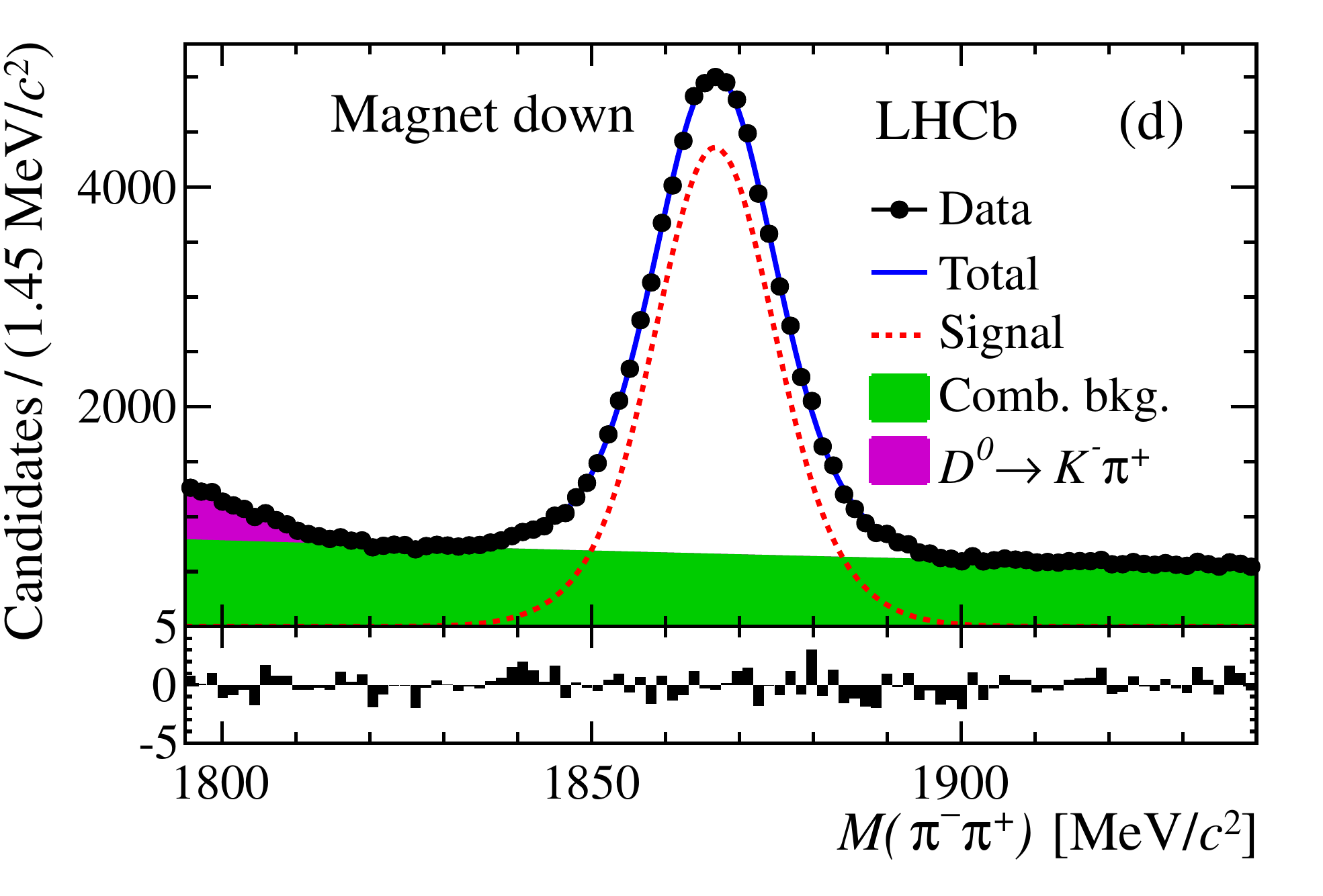}
\caption{Invariant mass distributions for $D^0\to K^-K^+$ (a,c) and $D^0\to \pi^-\pi^+$ 
(b,d) muon tagged candidates for the two magnet polarities. The fit result is superimposed,
with teh contributions from signal, combinatorial background and the $D^0\to K^-\pi^+$
reflection. Underneath each plot the pull in each mass bin is shown.}
\label{dacpB}
\end{figure}

The invariant mass distributions of muon tagged $D^0$ candidates are shown in Fig. \ref{dacpB}.
The signal yields are determined by a binned maximum likelihood fit. The signal is modelled
by a sum of two Gaussian functions with common means and different widths. The combinatorial 
background is represented by an exponential function. A Gaussian function is used to represent
the tail of the $D^0\to K^-\pi^+$ background. The total number of signal events is
($558.9\pm0.9$)$\times 10^3$ for $D^0\to K^-K^+$ and ($221.6\pm0.8$)$\times 10^3$ for 
$D^0\to \pi^-\pi^+$.

\subsection{Determination of the asymmetries}

The raw asymmetries are determined with simultaneous fits to the $D^0$ mass
distributions for positive and negative muon tags. A weighting procedure is used
in order to eliminate small differences in kinematic distributions between the two final 
states, that are result of  the particle identification, the phase space differences
and the correlation between the muons and the $D^0$. The event weighting 
improves the cancellation in eq. \ref{dacpsemi}. 

In some cases the $D^0$ flavor is not correctly tagged by the muon charge.
The mistag probability is estimated using the $D^0 \to K^-\pi^+$ decay. The
wrongly tagged decays include a small fraction of the doubly-Cabibbo-suppressed
decay $D^0 \to K^+\pi^-$, (0.393$\pm$0.007)\% \cite{ddbarmix}. After correcting for
this fraction the difference between  mistag probabilities for $D^0$ and 
$\overline{D}^0$ is (0.006$\pm$0.021)\% and is neglected.

Systematic uncertainties are due to:
\begin{itemize}
\item possible difference in the relative contribution 
from $B^0$ and $B^+$ decays between the $D^0 \to K^-K^+$ and $D^0 \to \pi^+\pi^-$ modes;
\item difference in $B$ decay time acceptance for the two $D^0$ modes;
\item differences in mistag rate for the two $D^0$ modes;
\item the $D^0$ fit model;
\item $\Lambda_c^+$ background in  $D^0 \to K^-K^+$;
\item low lifetime background in $D^0 \to \pi^+\pi^-$.
\end{itemize}

The small lifetime background in $D^0 \to \pi^+\pi^-$ is the dominant source of
systematic uncertainty. Many cross checks have been performed to verify the stability of the 
result, including the use of fiducial cuts on the muon,
a comparison between different trigger decisions and different particle identification
requirements. The stability of the raw asymmetries is also investigated as a function of
quantities such as the  $D^0$ decay time, the $b$-hadron flight distance, the
reconstructed $D^0$-$\mu$ invariant mass, the transverse momentum and pseudorapidity of the 
muon and the $D^0$ meson.

The difference in {\em CP} asymmetries between $D^0 \to K^-K^+$ and $D^0 \to \pi^+\pi^-$
modes, measured using $D^0$ mesons produced in semileptonic $B$ decays is found to be

\[
\Delta A_{CP} = (0.49\pm0.33(\mathrm{stat.})\pm0.14(\mathrm{syst.}))\%
\]

\section{Search for CPV in charged $D$ decays}

\subsection{Analysis strategy}

Searches for {\em CP} violation in charm have been mostly performed using
decays of the $D^0$ meson. A comprehensive programme, however, has to include
{\em CP} violation searches using charged $D$ decays as well. Since
there is no mixing in charged $D$, a non-zero {\em CP} asymmetry would
be an indication of direct CPV.

The large branching fraction of $D^0\to K^-K^+$ decay
compared to $D^0\to \pi^-\pi^+$, and of the
$D^+\to K^-K^+\pi^+$ decay compared to $D^+\to \pi^-\pi^+\pi^+$,
suggests a significant contribution from penguin amplitudes.

An investigation of CPV in the $\phi\pi^+$ region of the Dalitz plot
of the $D^+\to K^-K^+\pi^+$ decay --- hereafter referred to as
$D^+\to \phi\pi^+$ ---  is performed using the full 2011 LHCb data set 
(1.0 fb$^{-1}$). The $\phi\pi^+$ region is defined by 
$1.00< m_{K^-K^+}<1.04$ GeV/$c^2$.

The decay $D^+ \to K_s^0\pi^+$
is used as a control channel (CPV in this channel is possible via 
interference between Cabibbo-favoured and doubly Cabibbo-suppressed
amplitudes, but it is assumed to be negligible).

The {\em CP} asymmetry in the  $D^+ \to \phi\pi^+$ decay is, to first order, 
given by

\begin{equation}
A_{CP}(D^+ \to \phi\pi^+) = A_{\mathrm{raw}} (D^+ \to \phi\pi^+)
- A_{\mathrm{raw}} (D^+ \to K_s^0\pi^+) + A_{CP} (K^0/\overline{K}^0),
\end{equation}
where $A_{CP} (K^0/\overline{K}^0)$ is the correction for CPV in the
neutral kaon system and $A_{\mathrm{raw}}$ is defined as 

\[
A_{\mathrm{raw}} = \frac{N_{D^+} - N_{D^-}}{N_{D^+} + N_{D^-}}.
\]

A concurrent measurement of {\em CP} asymmetry is performed with the 
$D^+_s \to K_s^0\pi^+$ decay, using the $D^+_s \to \phi\pi^+$ decay
as a control channel. To first order one has

\begin{equation}
A_{CP}(D^+_s \to K_s^0 \pi^+) = A_{\mathrm{raw}} (D^+_s \to K_s^0\pi^+)
- A_{\mathrm{raw}} (D^+_s \to \phi\pi^+) + A_{CP} (K^0/\overline{K}^0).
\end{equation}

In both measurements it is assumed that asymmetries induced by the production 
and detection of the $D^+$ and the $D^-$ cancel in the difference between
signal and control channel raw asymmetries.

\begin{figure}[htb]
\centering
\includegraphics[width=7cm]{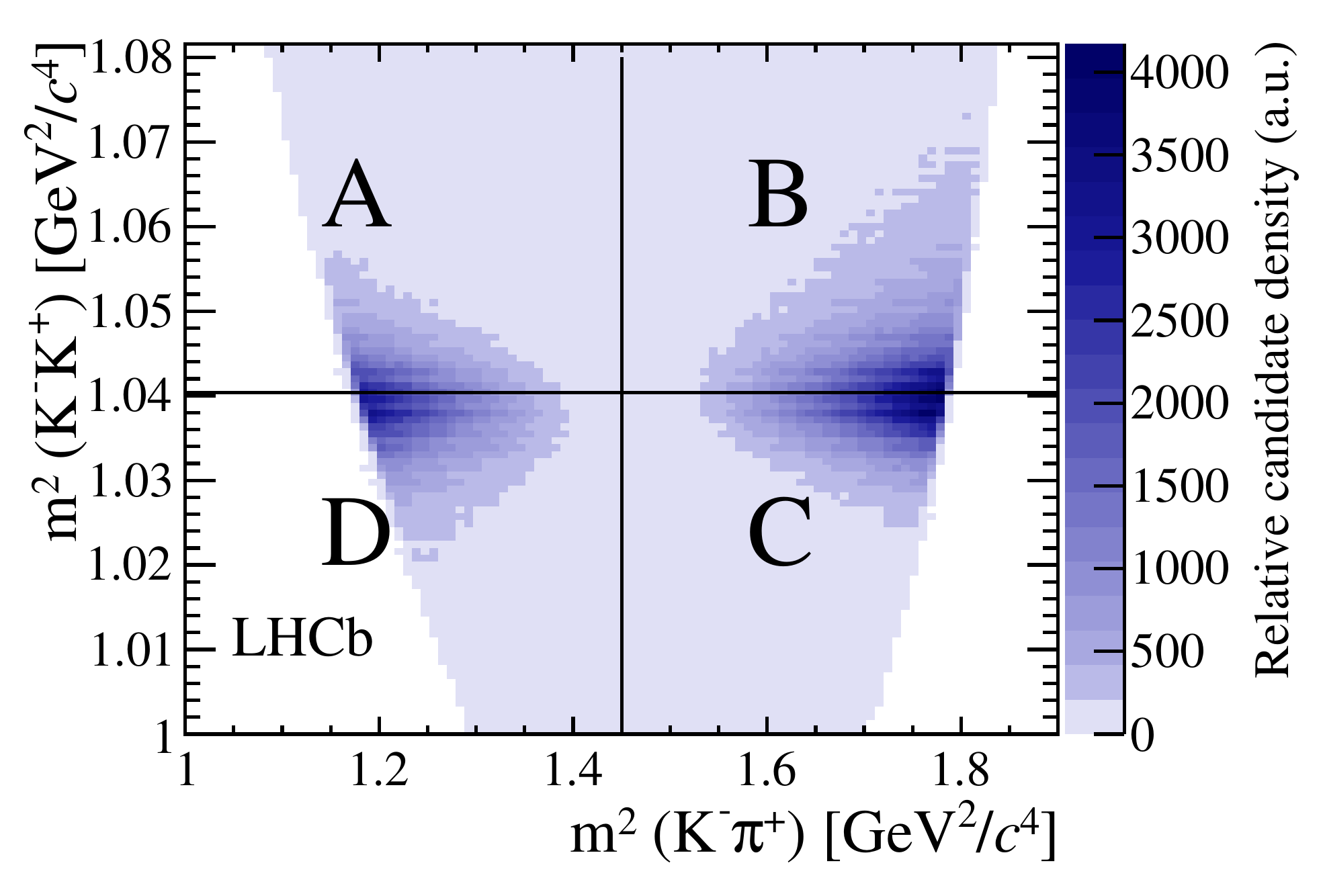}
\caption{Observed density of events in the Dalitz plot of the $D^+\to K^-K^+\pi^+$ decay.}
\label{hamish3b}
\end{figure}

Finally, a third measurement is performed. It is possible that
a constant {\em CP}-violating asymmetry is modulated by the rapidly varying
strong phase across the $\phi\pi^+$ region. In this case there could be
a cancellation of the asymmetry when different parts of the $\phi\pi^+$ region
are added to compute $A_{CP}$. The $\phi\pi^+$ region is divided into four
rectangular regions, A-D, shown in Fig. \ref{hamish3b}. A complementary observable 
is defined as

\begin{equation}
A_{CP|S} = \frac{1}{2} (A_{\mathrm{raw}}^A + A_{\mathrm{raw}}^C -
A_{\mathrm{raw}}^B + A_{\mathrm{raw}}^D).
\end{equation}

This observable is robust against systematic biases from the detector and is 
not affected by the $D^+$ production asymmetry.

\subsection{Data set and selection}

The $D^+_{(s)} \to \phi \pi^+$ candidates are reconstructed by combining
two oppositely charged particles identified as kaons by the RICH detector.
Pairs of oppositely charged pions are combined to form $K^0_s$ from
$D^+_{(s)} \to K_s^0 \pi^+$. Both $K^0_s$ and $D^+_{(s)}$ are required to
have a vertex with good fit quality. Further requirements are applied in
order to reduce the background from random track combinations, from partially
reconstructed charm decays and to avoid $D$ candidates coming from $B$ decays.
A detailed description of the selection criteria can be found in \cite{hamish}.

\begin{figure}[htb]
\centering
\includegraphics[width=7cm]{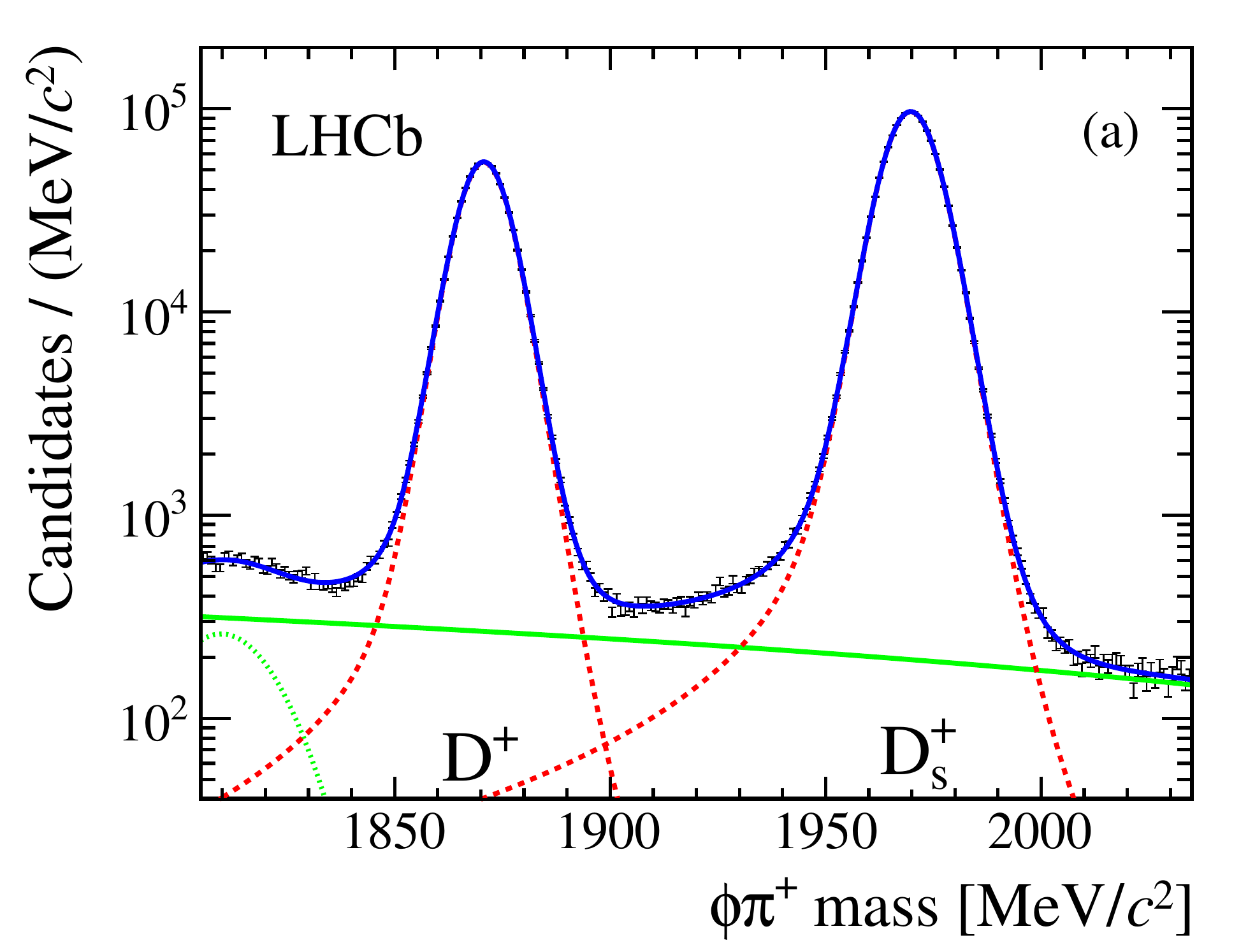}
\includegraphics[width=7cm]{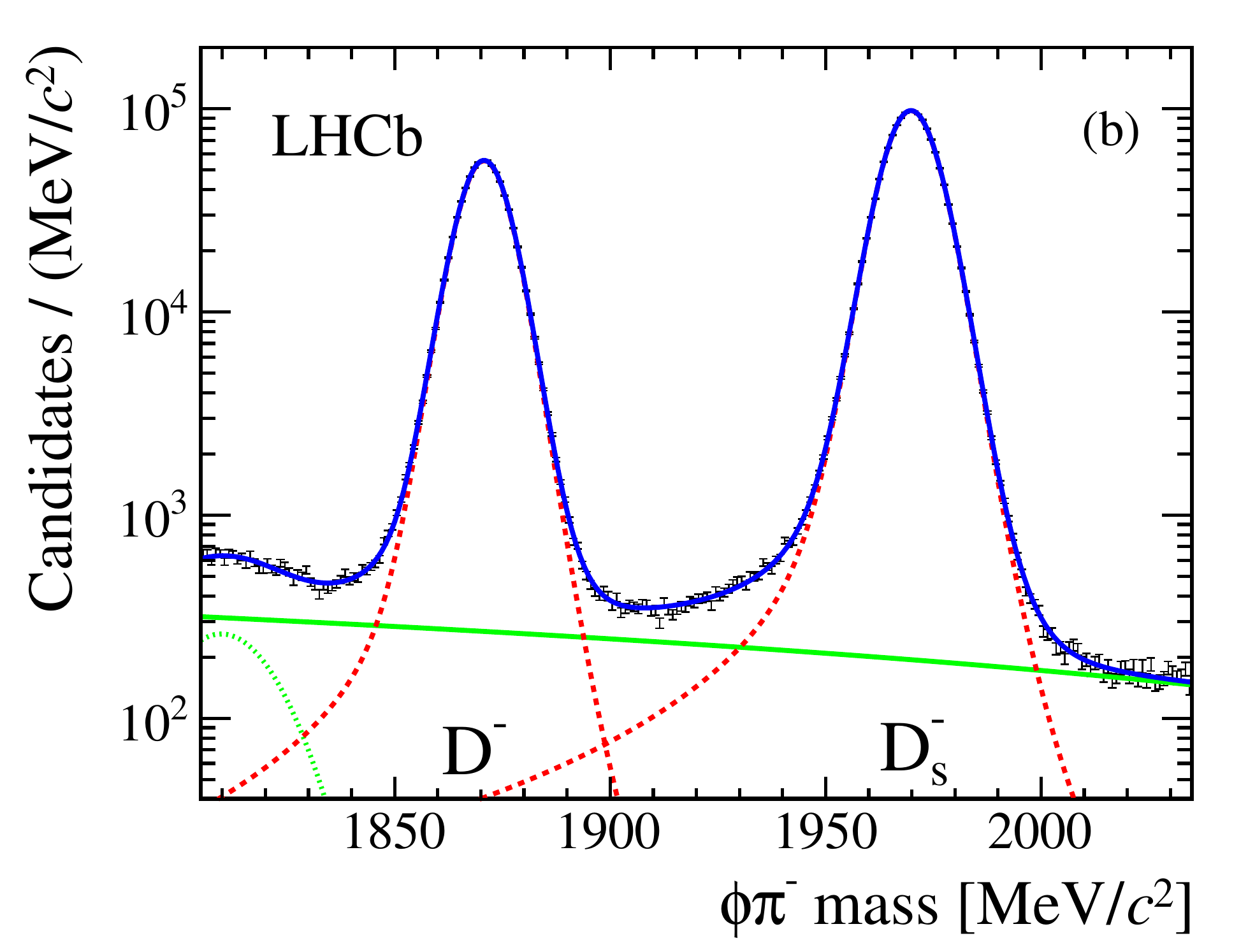}

\includegraphics[width=7cm]{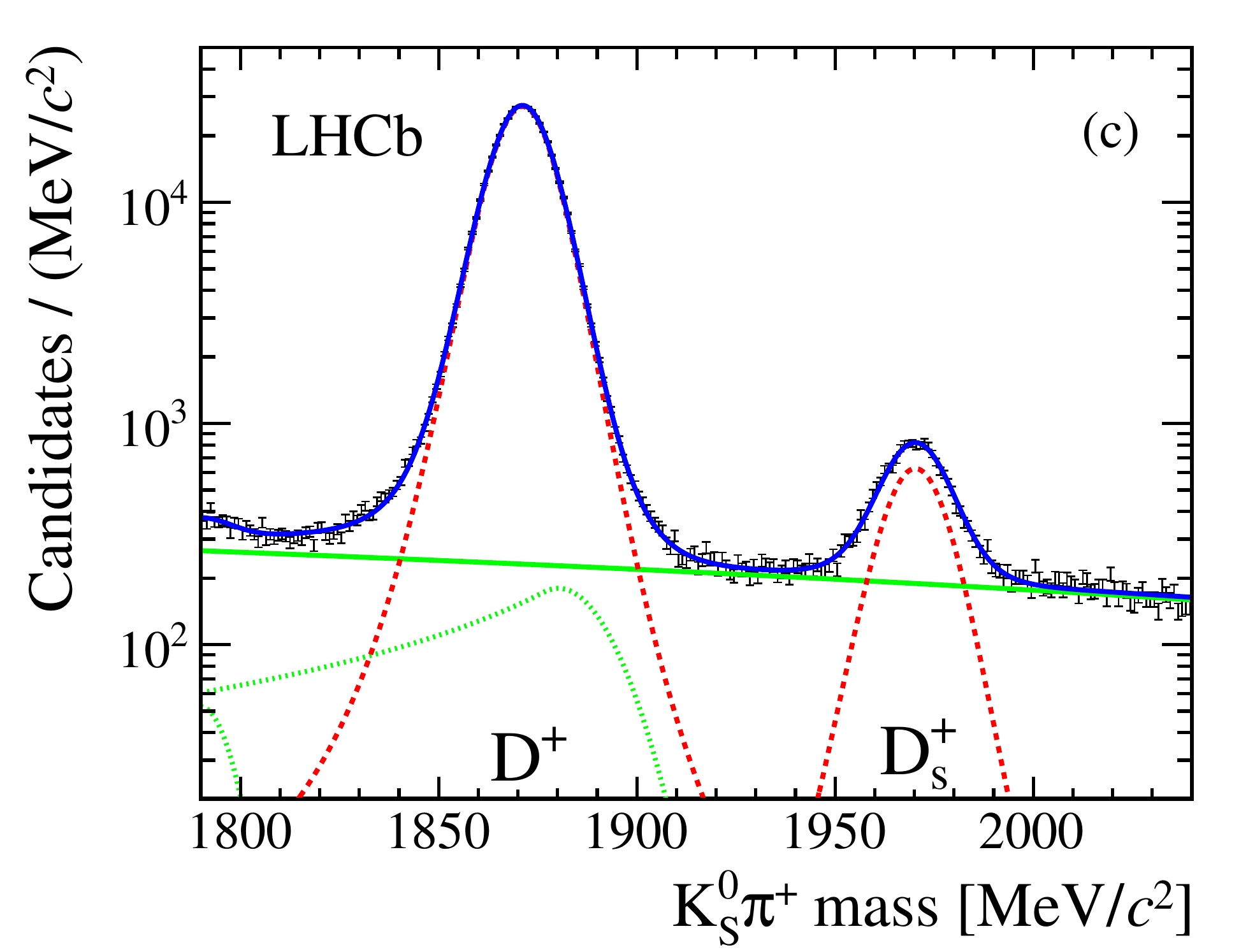}
\includegraphics[width=7cm]{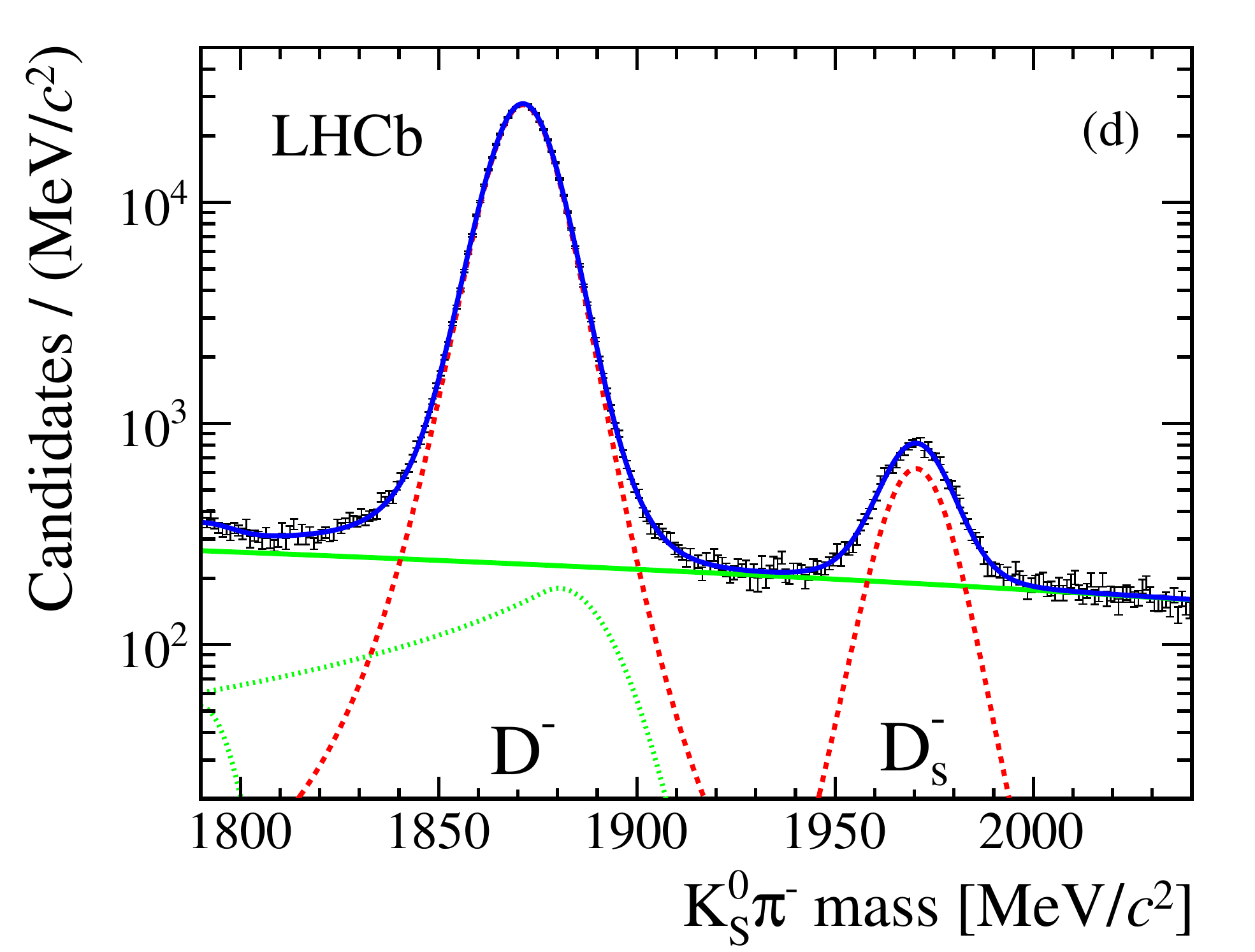}
\caption{Invariant mass distributions for (a) $D^+\to \phi\pi^+$, (b) $D^-\to \phi\pi^-$, 
(c) $D^+\to K_s^0\pi^+$ and (d) $D^-\to K_s^0\pi^-$ candidates. The data are represented
by symbols with error bars. The red dashed lines indicate the signal lineshapes. The green
solid line represents the combinatorial background whereas the dotted lines represent
the background from mis-reconstructed $D^+_s \to \phi\pi^+\pi^0$ or $D^+_s \to K_s^0K^+$.}
\label{hamish1}
\end{figure}

The invariant mass distributions of selected candidates are shown in
Fig. \ref{hamish1}.

\subsection{Determination of yields and asymmetries}

For the measurement of $A_{CP}$, the data was divided into 12 bins of $D$ transverse momentum $p_T$ and pseudorapidity $\eta$. The signal yields and the corresponding 
$A_{CP}$ are measured in each bin. A weighted average over the bins is performed to
obtain the final result. This procedure is adopted to reduce biases due to small 
differences in kinematics of the $D^+ \to \phi\pi^+$ and  $D^+ \to K_s^0\pi^+$
decays.

The shapes of the $D^+_{(s)} \to K_s^0\pi^+$ signals are described by single Cruijff
functions \cite{cruijff}. In the $\phi\pi^+$ final state Crystal Ball functions
\cite{cb} are added to the Cruijff functions to account for the tails of the mass peaks.
Monte Carlo simulations are used to test the signal lineshapes. The background is
fitted with a straight line plus a Gaussian component to account for partially
reconstructed $D^+_s \to K_s^0\pi^+\pi^0$ and $D^+_s \to \phi\pi^+\pi^0$ decays.
In the $K_s^0\pi^+$ case there is also a cross-feed component from the $D^+_s \to K_s^0K^+$. 

In each bin the data is further divided into four sets, according to the charge
of the $D$ and magnet polarity. The yields are determined by simultaneous fits
over the four sub-samples. The overall yields are shown in Table \ref{yields}.

\begin{table}[t]
\begin{center}
\begin{tabular}{cc}  
    Decay   &  yield ($\times 10^3$)   \\ \hline
 $D^+   \to \phi\pi^+$   & 1576.9 $\pm$ 1.5         \\
 $D^+_s \to \phi\pi^+$   & 3010.2 $\pm$ 2.2         \\  
 $D^+   \to K_s^0\pi^+$  & 1057.8 $\pm$ 1.2          \\      
 $D^+_s \to K_s^0\pi^+$  &   25.6 $\pm$ 0.2          \\ \hline
\end{tabular}
\caption{$D^+$ and $D^+_s$ yields from fits to the $\phi\pi^+$ and $K_s^0\pi^+$ spectra.}
\label{yields}
\end{center}
\end{table}

The main systematic uncertainties in $A_{CP}$ result from differences in the kinematics
of $\phi\pi^+$ and $K^0_s\pi^+$ final states, causing imperfect cancellation of
detector induced asymmetries, and from the response of the hardware trigger, which
is known to be charge asymmetric. Systematic effects from binning, fit model, kaon
detection asymmetry, $D$ mesons from $B$ decays, and from kaon {\em CP} violation are
also considered.

The results are 

\begin{eqnarray}
A_{CP} (D^+\phi\pi^+)&=&(-0.04\pm0.14\pm0.14)\%\nonumber \\
A_{CP|S} (D^+\phi\pi^+)&=&(-0.18\pm0.17\pm0.18)\%\nonumber \\
A_{CP} (D^+_s \to K_s^0\pi^+)&=&(+0.61\pm0.83\pm0.14)\%,\nonumber
\end{eqnarray} 
which are consistent with no {\em CP} violation.

\section{Concluding remarks}

CPV in charm is a phenomenon predicted by the SM  that was not yet observed. 
Finding CPV in charm is obviously a very imoportant goal, but understanding its 
nature will require different types of inputs.

The LHCb experiment has an extensive programme of CPV searches in the charm sector,
including two-, three- and four-body decays of both neutral and charged $D$ mesons.
With the 2011 data set (in 2012 LHCb recorded twice as much data at higher energy: 
2fb$^{-1}$ at 8 TeV), errors on the {\em CP} asymmetry are smaller than 0.2\%. This
means LHCb is now probing the regime of the SM expectations.

A naive combination of the two $\Delta A_{CP}$ measurements in two-body $D^0$ decays,
assuming negligible indirect CPV, yields

\[
\Delta A_{CP} = (-0.15\pm0.16)\%
\]

These and the results with charged $D$ mesons are the most sensitive searches and
show a consistent picture: at this time we have no evidence of CPV in the charm sector.



\Acknowledgements
I would like to thank the organizers of FPCP2013 for the excellent conference,
and the CNPq for the financial support.

\end{document}

%% file: econfmacros.tex
\def\beq{\begin{equation}}
\def\eeq#1{\label{#1}\end{equation}}
\def\eeqn{\end{equation}}

\def\beqa{\begin{eqnarray}}
\def\eeqa#1{\label{#1}\end{eqnarray}}
\def\eeqan{\end{eqnarray}}

\let\bar=\overbar

\def\Dslash{\not{\hbox{\kern-4pt $D$}}}
\def\dslash{\not{\hbox{\kern-2pt $\del$}}}

\def\msb{{\bar{\ssstyle M \kern -1pt S}}}

